\documentclass[11pt,a4paper]{article}
\usepackage{amsmath} 
\usepackage{amssymb} 
\usepackage{bm}  
\usepackage{color}

\def\onehalf{\textstyle{\frac{1}{2}}}

\def\be{\begin{equation}}
\def\ee{\end{equation}}
\def\ba{\begin{eqnarray}}
\def\ea{\end{eqnarray}}

\begin{document}

\begin{center}
{\Large \bf de Sitter relativity: a natural scenario for an evolving $\Lambda$}
\vskip 0.5cm

J. P. Beltr\'an Almeida$^{1}$, C. S. O. Mayor$^{2}$ and J. G. Pereira$^{2}$
\vskip 0.2cm
$^1${\it Departamento de F\'{\i}sica, Facultad de Ciencias B\'asicas \\
Universidad Antonio Nari\~no, Cra 3 Este 47A-15, Bogot\'a DC, Colombia}
\vskip 0.2cm
$^2${\it Instituto de F\'{\i}sica Te\'orica, UNESP-Univ Estadual Paulista \\
Caixa Postal 70532-2, 01156-970 S\~ao Paulo, Brazil}

\end{center}

\vskip 0.3cm
\begin{quote}
{\bf Abstract.}~{\footnotesize The dispersion relation of de Sitter special relativity is obtained in a simple and compact form, which is formally similar to the dispersion relation of ordinary special relativity. It is manifestly invariant under change of scale of mass, energy and momentum, and can thus be applied at any energy scale. When applied to the universe as a whole, the de Sitter special relativity is found to provide a natural scenario for the existence of an evolving cosmological term, and agrees in particular with the present-day observed value. It is furthermore consistent with a conformal cyclic view of the universe, in which the transition between two consecutive eras occurs through a conformal invariant spacetime.

}
\end{quote}


\section{Introduction}
\label{sec:intro}

Cosmological observations and quantum gravity arguments seem to indicate the necessity of an invariant length-parameter on both scales \cite{grf}. At the cosmological level, this parameter is related to the observed accelerated expansion rate of the universe \cite{obs}. At the quantum gravity level, this parameter is related to the Planck length $l_P$, which shows up as the threshold of quantum gravity~\cite{maj-smo-ame}. A natural question then arises: could these two length-parameters have one and the same origin? To answer this question, let us go back to special relativity. As is well-known, time and space are not absolute: seen from different inertial frames, time dilates and space contracts. Of course, no one doubts these facts are correct, at least in the energy scales we are used to. However, when one considers the, by now unreachable but conceptually well defined Planck scale, they conflict with the existence of invariant parameters that show up at that scale. For example, considering that the Planck length
\be
l_P = \left(\frac{G \hbar}{c^3} \right)^{1/2}
\ee
depends only on fundamental constants, it cannot change under Lorentz transformations. The only way to circumvent this conundrum is to introduce a new special relativity. One of the first attempts in this direction is the so-called doubly special relativity \cite{dsr}, which is based on the $\kappa$-deformed Poincar\'e group. This model introduces changes in the special relativity dispersion relation\footnote{ We are going to use the first half of the Greek alphabet ($\alpha, \beta, \gamma, \dots$) to denote indices related to Minkowski spacetime, and the second half ($\mu, \nu, \rho, \dots$) to denote indices related to de Sitter spacetime, as well as to any spacetime that reduces locally to de Sitter.}
\be
\eta^{\alpha \beta} \, p_\alpha p_\beta = m^2 c^2,
\label{SRDisRel}
\ee
with $p_\alpha = m c u_\alpha$, in such a way to allow the existence of a Lorentz-invariant length at the Planck scale. More specifically, in order to allow the presence of an energy scale, scale-suppressed terms of higher order in the momentum are introduced. Because these terms are not invariant under change of scale of mass, energy and momentum, this kind of models can be applied at the Planck scale only.

A different approach to this problem shows up from the trivial observation that Lorentz transformations do not change the vanishing curvature of the Minkowski spacetime. Considering that the scalar curvature $R$ of any homogeneous space, including Minkowski, is of the form
\[
R \sim \pm \, l^{-2},
\]
with $l$ the pseudo-radius of the homogeneous space, we can say that Lorentz transformations preserve the infinite pseudo-radius of Minkowski spacetime. Now, Lorentz transformation can be defined in any isotropic and homogeneous spacetime \cite{Jackson}. This means that, in addition to Minkowski, they can also be defined in both de Sitter and anti-de Sitter spaces, the only four-dimensional pseudo-Euclidian homogeneous spacetimes \cite{weinberg}. This is clear from the fact that Lorentz is a subgroup of both de Sitter and anti-de Sitter groups. Then comes the point: like in Minkowski, Lorentz transformations do not change the scalar curvature, and consequently leave the de Sitter pseudo-radius $l$ invariant. Although somewhat hidden in Minkowski, because what is preserved in that case is an infinite parameter, in the de Sitter and anti-de Sitter spaces this property becomes clearly manifest. {\em If the Planck length is to be invariant under Lorentz transformations, therefore, it must represent the pseudo-radius of spacetime at the Planck scale, which must then be either a de Sitter or an anti-de Sitter space}. From now on, for the sake of definiteness, we are going to consider only the de Sitter case.

Now, if spacetime is not Minkowski, but de Sitter, instead of ruled by Poincar\'e, spacetime kinematics will be ruled by the de Sitter group. This means that ordinary special relativity must be replaced by de Sitter special relativity \cite{paper1,Guo,Cacciatori:2007in}. In this theory, even though the spatial isotropy and the relation between inertia-free frames are still governed by the Lorentz group, it involves an invariant length-parameter related to the de Sitter curvature. A crucial difference in relation to the doubly special relativity is that the de Sitter dispersion relation is invariant under change of scale of mass, energy and momentum, and can consequently be applied at any energy scale, from quantum gravity to cosmology. Concerning the specific case of cosmology, the de Sitter special relativity naturally introduces a pervading cosmological term $\Lambda$, which appears as a purely kinematic effect. Considering that dark energy, the entity responsible for the observed acceleration of the universe expansion rate, seems to be in fact a cosmological term \cite{Lambda}, the de Sitter relativity can be considered a natural model to approach the universe acceleration expansion.

Motivated by the above arguments, the purpose of this paper is to explore further some specific properties of the de Sitter special relativity. First, we are going to show that the dispersion relation of this theory can be written in a simple and compact form, which is formally similar to the dispersion relation of ordinary special relativity. The only differences are the notions of energy and momentum, whose expressions include now a conformal component, and the mass term, which acquires an additional term related to the spin of the particle. We then explore some cosmological consequences of these changes. In particular, we are going to show that, when applied to the universe as a whole, the de Sitter special relativity provides a natural scenario for the existence of an evolving cosmological term. It is consistent with a very high primordial $\Lambda$, as well as with its currently observed value. Finally, it is discussed why the de Sitter special relativity points to a cyclic view of the universe, where the remote future of one era becomes the initial condition of the next one.

\section{Basics of de Sitter special relativity}

For the sake of completeness, we review in this section the fundamentals of the de Sitter special relativity. In particular, we discuss how the de Sitter transitivity properties change the notions of energy and momentum change, as well as analyze the concomitant changes that occur in the corresponding conservation laws.

\subsection{The de Sitter spacetime}

The de Sitter spacetime, denoted $dS(4,1)$, is a hyperbolic, maximally symmetric curved spacetime \cite{ellis}. It can be defined as a hyper-surface in the ``host'' pseudo-Euclidean spaces ${\bf E}^{4,1}$, inclusion whose points in Cartesian coordinates $\chi^A$ $(A, B, ... = 0, ..., 4)$ satisfy
\[
\eta_{AB} \chi^A \chi^B = - \, l^2,
\]
where $\eta_{AB}$ = $(+1,-1,-1,-1,-1)$ and $l$ is the de Sitter length-parameter (or pseudo-radius). Its kinematic group is the pseudo-orthogonal group $SO(4,1)$, and it is defined as the quotient space
\[
dS(4,1) = SO(4,1)/ {\mathcal L},
\]
with ${\mathcal L}$ the Lorentz group. The stereographic coordinates are obtained through a projection of the hyperboloid onto a target Minkowski spacetime with coordinates $x^\alpha$ and metric $\eta_{\alpha \beta}$. They are defined by \cite{gursey}
\be
\chi^\alpha = \Omega(x) \, x^\alpha \qquad \mbox{and} \qquad
\chi^4 = -\, \Omega(x) \left(1 + \frac{\sigma^2}{4 l^2} \right),
\label{stepro}
\ee 
where
\be
\Omega(x) = \frac{1}{1 - {\sigma^2}/{4 l^2}},
\label{n}
\ee
with $\sigma^2$ the Lorentz-invariant quadratic form $\sigma^2 = \eta_{\alpha \beta} \, x^\alpha x^\beta$.

\subsection{de Sitter generators}

The generators of infinitesimal de Sitter transformations are
\be
\hat {\mathcal L}_{A B} = \eta_{AC} \, \chi^C \, \frac{\partial}{\partial \chi^B} -
\eta_{BC} \, \chi^C \, \frac{\partial}{\partial \chi^A}.
\label{dsgene}
\ee
They can be decomposed according to
\be\label{decomp}
\hat  L_{\alpha \beta} = \hat {\mathcal L}_{\alpha \beta} \qquad \mbox{and} \qquad
\hat \Pi_\alpha = \frac{1}{l} \hat {\mathcal L}_{\alpha 4}
\ee
where
\be\label{lorentzdS}
\hat L_{\alpha \beta} =
\eta_{\alpha \gamma} \, x^\gamma \, \hat P_\beta -
\eta_{\beta \gamma} \, x^\gamma \, \hat P_\alpha
\ee
are the Lorentz generators, and
\be\label{pi}
\hat \Pi_\alpha \equiv \frac{\hat L_{\alpha 4}}{l} =
\hat P_\alpha - \frac{1}{4 l^2} \, \hat K_\alpha
\ee
are the so-called de Sitter ``translation'' generators, with
\be
\hat P_\alpha = \partial/ \partial x^\alpha \quad \mbox{and} \quad
\hat K_\alpha = \left(2 \eta_{\alpha \gamma} \, x^\gamma x^\delta - \sigma^2 \delta_{\alpha}{}^{\delta} \right) 
\partial/ \partial x^\delta,
\label{TransGenerators}
\ee
respectively, the translation and the proper conformal generators \cite{livro}. For $l \to \infty$, they reduce to the generators of the Poincar\'e group, the kinematic group of Minkowski spacetime. We see from the generators (\ref{lorentzdS}) and (\ref{pi}) that, from the algebraic point of view, the only difference between Minkowski and de Sitter spacetimes is the notion of uniformity. Whereas in Minkowski the uniformity is defined by ordinary translations, in de Sitter it is defined by a combination of translation and proper conformal transformation.

\subsection{Killing vectors}
\label{Killing}

The metric of the de Sitter spacetime is obtained from
\be
g_{\mu \nu} = h^\alpha_\mu \, h^\beta_\nu \; \eta_{\alpha \beta},
\label{dSmetric1}
\ee
where
\be
h^\alpha_\mu = \Omega(x) \, \delta^\alpha_\mu \qquad
h_\alpha^\mu = \Omega^{-1}(x) \, \delta_\alpha^\mu
\ee
is the de Sitter tetrad, with $\delta^\alpha_\mu$ a trivial tetrad. In terms of the de Sitter coordinates $x^\mu$, it assumes the conformally flat form
\be
g_{\mu \nu} = \Omega^2(x) \, \eta_{\mu \nu}.
\label{44}
\ee

An infinitesimal de Sitter transformation is written as
\be\label{ds-t}
\delta x^\mu \equiv \delta_{L} x^\mu + \delta_{\Pi} x^\mu = 
\onehalf \epsilon^{\alpha \beta} \hat{L}_{\alpha \beta} x^\mu -
\epsilon^\alpha \hat{\Pi}_\alpha x^\mu,
\ee
where $\epsilon^{\alpha \beta} = - \epsilon^{\beta \alpha}$ {and} $\epsilon^{\alpha}$ are the transformation parameters. Substituting the generators (\ref{lorentzdS}) and (\ref{pi}), the Lorentz transformations can be written as
\be
\delta_{L} x^\mu = - \xi^{\;\mu}_{\alpha \beta} \, \epsilon^{\alpha \beta},
\ee
where
\be
\xi^{\;\mu}_{\alpha \beta} = \left( \eta_{\alpha \gamma} \, \delta^\mu_\beta - 
\eta_{\beta \gamma} \, \delta^\mu_\alpha \right) x^\gamma
\ee
are the corresponding Killing vectors, whereas the de Sitter ``translations'' become
\be
\delta_{\Pi} x^\mu = - \xi^{\, \mu}_{\alpha} \, \epsilon^{\alpha},
\label{dStrans}
\ee
where
\be
\xi^{\, \mu}_{\alpha} = \delta^\mu_\alpha - \frac{1}{4l^2} \bar{\delta}_\alpha^\mu
\label{dsKilling}
\ee
are the correspponding Killing vectors, with
\be
\bar{\delta}_\alpha^\mu = 2 \delta^{\mu}_\gamma \, x^\gamma x_\alpha -
\sigma^2 \delta_\alpha^\mu
\label{deltabar}
\ee
a conformal generalization of the Kronecker delta.

\subsection{Conserved quantities}
\label{conserlaw}

Let us consider the infinitesimal de Sitter ``translation'' $\delta_{\Pi} x^\mu$ defined in Eq.~(\ref{dStrans}). Under such transformation, the metric changes according to
\be
\delta g_{\mu \nu} = - \, \xi^{\alpha}_{\mu} \, \nabla_\nu \epsilon_\alpha -
\, \xi^{\alpha}_{\nu} \, \nabla_\mu \epsilon_\alpha,
\label{metricTrans}
\ee
where we have already used that $\xi^{\alpha}_{\mu}$ satisfy the Killing equation
\be
\nabla_\nu \xi^{\alpha}_{\mu} + \nabla_\mu \xi^{\alpha}_{\nu} = 0.
\ee
From Noether's theorem, the invariance of any matter Lagrangian ${\mathcal L}_m$ under this transformation yields the conservation law
\be
\nabla_\mu \Pi^{\rho \mu} = 0,
\label{dSconservation}
\ee
where
\be
\Pi^{\rho \mu} \equiv \xi^{\rho}_{\alpha} \, T^{\alpha \mu} =
T^{\rho \mu} - \frac{1}{4l^2} \, {K}^{\rho \mu},
\label{TmK}
\ee
with $T^{\rho \mu}$ the symmetric energy-momentum current, and $K^{\rho \mu}$ the proper conformal current~\cite{coleman}
\be
K^{\rho \mu} \equiv \bar{\delta}_\alpha^\rho \, T^{\alpha \mu} =
\left(2 \delta^{\rho}_\gamma \, x^\gamma x_\alpha -
\sigma^2 \delta_\alpha^\rho \right) T^{\alpha \mu}.
\label{KdelT}
\ee
We see in this way that, due to the changes in the spacetime uniformity, instead of the energy-momentum current, the conserved quantity is a combination of the energy-momentum $T^{\rho \mu}$ and the proper conformal current $K^{\rho \mu}$. Of course, the notions of energy and momentum will also change in de Sitter relativity. Energy, in particular, will have the form
\be
E \equiv \Pi^{00} = E_T - \frac{1}{4 l^2} E_K,
\ee
where $E_T$ is the ordinary translational energy, and $E_K$ is the proper conformal energy.

In general, neither $T^{\mu \nu}$ nor $K^{\mu \nu}$ is conserved separately. In fact, as an explicit calculation shows,
\be
\nabla_\mu T^{\mu \nu} = \frac{2 \, T^\rho{}_\rho \, x^\nu}{4l^2 - \sigma^2} \qquad \mbox{and} \qquad
\nabla_\mu {K}^{\mu \nu} = \frac{2 \, T^\rho{}_\rho \, x^\nu}{1 - \sigma^2/4l^2}.
\label{Conser1}
\ee
Only for matter with traceless energy--momentum tensor the currents $T^{\mu \nu}$ and ${K}^{\mu \nu}$ are separately conserved. In the formal limit $l \to \infty$, we obtain
\be
\nabla_\mu T^{\mu \nu} = 0 \qquad \mbox{and} \qquad
\nabla_\mu K^{\mu \nu} = 2 \, T^\rho{}_\rho \, x^\nu.
\label{Conser2}
\ee
On the other hand, in the formal limit $l \to 0$, we get
\be
\nabla_\mu T^{\mu \nu} = - \, 2 \, T^\rho{}_\rho \, \frac{x^\nu}{\sigma^2} \qquad \mbox{and} \qquad
\nabla_\mu {K}^{\mu \nu} = 0.
\label{Conser3}
\ee
In this limit, physics becomes conformally invariant, and the proper conformal current turns out to be conserved.

\subsection{Casimir Invariant and dispersion relation}

The first Casimir invariant of the de Sitter group is \cite{gursey}
\be
\hat{\mathcal C}_{dS} = - \frac{1}{2 l^2} \, \eta^{AC} \, \eta^{BD} \, \hat{J}_{AB} \, \hat{J}_{CD},
\label{1dsCas0}
\ee
where
\be
\hat{J}_{AB} = \hat{\mathcal L}_{AB} + \hat{\mathcal S}_{AB},
\ee
with $\hat{\mathcal L}_{AB}$ the orbital generators and $\hat{\mathcal S}_{AB}$ the matrix spin generators. For the case of a spin-0 particle, it reduces to
\be
\hat {\mathcal C}_{dS} = - \frac{1}{2 l^2} \, \eta^{AC} \, \eta^{BD} \,\hat {\mathcal L}_{AB} \, \hat {\mathcal L}_{CD},
\label{1dsCas}
\ee
In terms of the generators $\hat{L}_{ab}$ and $\hat{\Pi}_a$, it assumes the form
\be
\hat {\mathcal C}_{dS} = \eta^{\alpha \beta} \, \hat \Pi_\alpha \, \hat \Pi_\beta -
\frac{1}{2 l^2} \, \eta^{\alpha \beta} \, \eta^{\gamma \delta} \, \hat L_{\alpha \gamma} \, \hat L_{\beta \delta}.
\label{cas1}
\ee
Considering again a particle of mass $m$, we get
\be
{\mathcal C}_{dS} = g_{\mu \nu} \, \pi^\mu \, \pi^\nu -
\frac{1}{2 l^2} \, g_{\mu \nu} \, g_{\rho \sigma} \, l^{\mu \rho} \, l^{\nu \sigma},
\label{cas1Bis}
\ee
where
\be
\pi^\mu = \xi_{\alpha}^{\mu} \, p^\alpha \qquad \mbox{and} \qquad 
l^{\mu \rho} = \xi_{\;\alpha}^{\mu \rho} \, p^\alpha
\ee
are respectively the de Sitter linear mo\-men\-tum and the angular momentum, with $p^\alpha = m c u^\alpha$ the ordinary (Poincar\'e) momentum. The de Sitter linear momentum is related to the de Sitter current through $\pi^\mu = \int d^3 x \, \Pi^{0 \mu}$.

Now, due to the fact that de Sitter spacetime is isotropic and homogeneous, the angular momentum $l^{\mu \rho}$ vanishes for free particles, that is, for particles moving along a geodesic of the de Sitter spacetime. This is similar to what happens in Minkowski spacetime, which is also isotropic and homogeneous. The Casimir invariant reduces then to the form
\be
{\mathcal C}_{dS} = g_{\mu \nu} \, \pi^\mu \pi^\nu.
\label{cas2}
\ee
For particles belonging to representations of the {\it principal series}, the eigenvalues of the Casimir operator are given by \cite{dix}
\begin{equation}
m^2 - \frac{\hbar^2}{c^2 l^2} \, \left[{\sf s}({\sf s}+1)-2 \right]
 \equiv \mu_{\sf s}^2,
\label{rforcs}
\end{equation}
with $m$ the mass and ${\sf s}$ the spin of the field under consideration. The eigenvalue $\mu$ is sometimes referred to as the de Sitter mass. Notice that for ${\sf s} = 1$, the de Sitter and the Poincar\'e masses coincide: $m = \mu_1$. For ${\sf s} = 0$, on the other hand, it assumes the form
\begin{equation}
m^2 + \frac{2 \hbar^2}{c^2 l^2} \equiv \mu_0^2.
\label{rforcs0}
\end{equation}
The dispersion relation of de Sitter special relativity is, consequently, given by
\be
g_{\mu \nu} \, \pi^\mu \pi^\nu = \mu_0^2 c^2.
\label{dsDisRel}
\ee
It is formally similar to the ordinary dispersion relation (\ref{SRDisRel}), but with two fundamental differences: the momentum and the mass are replaced by their de Sitter counterparts. Of course, for $l \to \infty$, it reduces to the ordinary special relativity dispersion relation. Notice furthermore that it is invariant under change of scale of mass, energy and momentum, which means that it can be applied at any scale, from quantum gravity to cosmology.

\section{Cosmological term as a kinematic effect}
\label{OriLambda}

The cosmological term as a kinematic effect related to the de Sitter group has already been considered in a previous publication \cite{paper2}. Considering that we have since then improved the formalism, as well as obtained a new covariant interpretation of the results in terms of co-moving frames, which allowed us to discuss the present cosmological application on more solid grounds, we readdress this discussion here. In ordinary general relativity, spacetime is assumed to reduce locally to Minkowski, whose kinematics is ruled by the Poincar\'e group, and whose homogeneity is related to the translational Killing vectors $\delta^\rho_\alpha$. As a consequence, Einstein equation is written in the form
\be
\delta^\rho_\alpha \, G^{\alpha \mu} = 
\frac{8 \pi G}{c^4} \, \delta^\rho_\alpha \, T^{\alpha \mu},
\ee
with $G^{\alpha \mu} = R^{\alpha \mu} - \frac{1}{2} h^{\alpha \mu} R$ the Einstein tensor. If instead of Poincar\'e, local kinematics is to be ruled by the de Sitter group, spacetime must reduce locally to de Sitter, whose homogeneity is defined by the ``translational'' Killing vectors $\xi^\alpha_{\mu}$. In this case, Einstein equation must be generalized to
\be
\xi^\rho_\alpha \, G^{\alpha \mu} = 
\frac{8 \pi G}{c^4} \, \xi^\rho_\alpha \, T^{\alpha \mu}.
\label{nee0}
\ee
Using the Killing vectors~(\ref{dsKilling}), as well as the definition~(\ref{TmK}), the field equation~(\ref{nee0}) can be rewritten in the form
\be
G_{T}^{\rho \mu} - \frac{1}{4l^2} \, G_K^{\rho \mu} =
\frac{8 \pi G}{c^4} \left( T^{\rho \mu} - \frac{1}{4l^2} \, K^{\rho \mu} \right),
\label{nee}
\ee
where we have introduced the notations
\be
G_{T}^{\rho \mu} = \delta^\rho_\alpha \, G^{\alpha \mu} \qquad \mbox{and} \qquad
G_{K}^{\rho \mu} = \bar{\delta}^\rho_\alpha \, G^{\alpha \mu}.
\label{35}
\ee
In the limit $l \to \infty$, it reduces to the usual Einstein equation, which is consistent with ordinary special relativity.

It is important to note that the replacement of ordinary (Poincar\'e-based) special relativity by de Sitter relativity changes only the local spacetime symmetry, not the dynamics of general relativity. As a matter of fact, it changes only the strong equivalence principle, which states now that in any locally inertial frame, the non-gravitational laws of physics are those of de Sitter special relativity. This means essentially that Einstein equation
\be
G_T^{\rho \mu} =
\frac{8 \pi G}{c^4} \, T^{\rho \mu},
\label{oee}
\ee
which describes the dynamic spacetime curvature, and the kinematic equation
\be
G_K^{\rho \mu} =
\frac{8 \pi G}{c^4}\, K^{\rho \mu},
\label{dSitterEffects}
\ee
which is an equation describing the de Sitter effects on the spacetime geometry, can be solved separately. Of course, the total spacetime curvature will always be composed of two parts: a dynamic part $G_T^{\rho \mu}$, whose source is the energy-momentum tensor of matter, and a kinematic part $G_K^{\rho \mu}$, whose source is the proper conformal current of matter. The importance of each part depends on the value of the pseudo-radius $l$.

Now, since in the present context $G_K^{\rho \mu}$ represents the curvature of the underlying de Sitter spacetime, it depends necessarily on $l$. As a matter of fact, if we recall that, for a de Sitter spacetime,
\be
G^{\alpha \mu} = \frac{3}{l^2} \, h^{\alpha \mu},
\label{dSGab}
\ee
and taking into account that $\Lambda = 3 / l^2$, with $\Lambda$ the cosmological term, we see from Eq.~(\ref{35}) that
\be
G^{\rho \mu} = \Lambda \, \bar{\delta}^\rho_\alpha \, h^{\alpha \mu}.
\ee
Substituting into Eq.~(\ref{dSitterEffects}), and taking the trace on both sides, we get
\be
\Lambda = \frac{4 \pi G}{c^4} \frac{K_\mu{}^{\mu}}{\sigma^2}.
\label{dSitterEffects3}
\ee
Using Eq.~(\ref{KdelT}), the cosmological term is then found to be given by
\be
\Lambda = \frac{4 \pi G}{c^4} \left(2 \, T_{\rho \mu} \, u^\rho u^\mu -
T^\mu{}_\mu \right),
\label{dSitterEffects6}
\ee
where
\be
u^\mu = \frac{x^\mu}{s}
\ee
is the four-velocity in {\em comoving coordinates}, with $s$ the invariant form $s = ({g_{\mu \nu} x^\mu x^\nu})^{1/2} \equiv \Omega(x) \, \sigma$. In fact, observe that the variation of the four-velocity of a given point of spacetime depends only on the change of the invariant form $s$. 

According to de Sitter special relativity, therefore, the source of the local de Sitter spacetime, or equivalently, of the local value of the cosmological term, is not the energy-momentum current, but the (trace of the) proper conformal current of ordinary matter. It should be noticed, however, that there is a crucial difference between the local de Sitter spacetime underlying all physical systems, and the usual notion of de Sitter spacetime: since equation~(\ref{dSitterEffects6}) is purely kinematics, and considering that the conformal current $K^{\rho \mu}$ vanishes outside the source, in this region the local de Sitter spacetime becomes the flat Minkowski space. This is quite similar to a proposal made by Mansouri in a different context \cite{mansouri}.

\section{The de Sitter scenario for a variable $\Lambda$}

As we have seen, what is conserved in de Sitter special relativity is a combination of energy-momentum and proper conformal currents. Considering that neither of these currents is individually conserved, energy-momentum can be transformed into proper conformal current, and vice-versa. Since the proper conformal current appears as the source of $\Lambda$, this means that ordinary matter can be transformed into dark energy, and vice versa. This mechanism provides a natural scenario for an evolving $\Lambda$. When applied to the universe as a whole, it predicts the existence of a pervading cosmological term whose value, at each time, is determined by the matter content of the universe.

Let us begin by considering a universe with an infinite cosmological term. Such universe is obtained in the limit $l \to 0$,\footnote{We remark that, in order to consider the limit $l \to 0$, a different parametrization of the generators must be used. In fact, instead of the parametrization (\ref{pi}), which is appropriate to study the limit $l \to \infty$, it would be necessary to use $\hat{\Pi}_\rho \equiv l \hat{L}_{\rho 4}$ \cite{paper1}.} under which spacetime becomes a flat, singular cone-spacetime, transitive under proper conformal transformations \cite{paper1}. Its kinematic group is the conformal Poincar\'e group, the semi-direct product between Lorentz and the proper conformal groups \cite{ConPoin}. In this limit, physics becomes conformal invariant, and the proper conformal current turns out to be conserved (see section~\ref{conserlaw}). Observe that, since this spacetime is transitive under proper conformal transformations only, the usual notions of space and time will not be present. As a consequence, the energy-momentum current cannot be defined, which means that the degrees of freedom of gravity are turned off. Only the degrees of freedom associated with the proper conformal transformations are active.\footnote{This is somewhat similar to the ``Weyl Curvature Hypothesis" of Penrose \cite{pen0}.} It is also interesting to observe that, under such extreme situation, the thermodynamic properties of the de Sitter horizon fit quite reasonably to what one would expect for an initial condition for the universe: it has an infinite temperature, vanishing entropy, and vanishing energy; the energy density, however, is infinite \cite{abap}.

It is important to remark that the limit $l \to 0$ is purely formal in the sense that quantum effects preclude it to be fully performed. The outcome of this limit, that is, the cone spacetime, must be thought of as the frozen geometrical structure behind the quantum spacetime fluctuations taking place at the Planck scale. A more realistic situation is to assume that the de Sitter pseudo-radius $l$ is of the order of the Planck length $l_P$, in which case the cosmological term would have the value
\be
\Lambda \simeq 1.2 \times 10^{66}\, {\rm cm}^{-2}.
\label{PlanLam}
\ee
Notice that, due to the fact that spacetime becomes locally a high-$\Lambda$ de Sitter spacetime, it will naturally be endowed with a causal de Sitter horizon, and consequently with a holographic structure \cite{holo}. As the $\Lambda$ term decays and the universe expands, the proper conformal current is gradually transformed into energy-momentum current, giving rise to a FRW universe. Concomitantly, spacetime becomes transitive under a combination of translations and proper conformal transformations. This means that our usual (translational) notion of space and time begin to emerge, though at this stage spacetime is still preponderantly transitive under proper conformal transformation.

As the proper conformal current is continuously transformed into energy-momentum current, the matter content of the universe turns out to be described by a sum of many different components. As a simple example, we consider here a one-component perfect fluid, whose energy-momentum tensor is of the form
\be
T^{\rho}{}_{\mu} = (\varepsilon + p) \, u^\rho u_\mu - p \, \delta^{\rho}{}_{\mu},
\ee
with $\varepsilon$ the energy density, and $p$ the pressure. Substituting in Eq.~(\ref{dSitterEffects6}), it yields the cosmological term
\be
\Lambda = \frac{4 \pi G}{c^4} \, (\varepsilon + 3 p).
\label{dSitterEffects7}
\ee
Observe that, even though $\Lambda$ can decay to small values, it cannot vanish.\footnote{It is interesting to note that, in order to have a vanishing $\Lambda$, it would be necessary to suppose the existence of an exotic fluid satisfying the unphysical equation of state $\varepsilon = - 3 p$.}
This means that, similar to the limit $l \to 0$, the limit $l \to \infty$ cannot be accomplished either because it would lead to an empty universe represented by Minkowski spacetime.
At the present time, the matter content of the universe can be accurately described by dust, which is characterized by $p = 0$. In this case, expression (\ref{dSitterEffects7}) yields
\be
\Lambda = \frac{4 \pi G}{c^4} \, \varepsilon.
\label{dSitterEffects8}
\ee
Using the current value for the energy density of the universe, we get
\be
\Lambda \simeq 10^{-56}~\mbox{cm}^{-2},
\label{obs}
\ee
which is consistent with recent observations~\cite{obs}. For such value of $\Lambda$, spacetime becomes preponderantly transitive under ordinary spacetime translations.

Now, as is well known, in order to allow the formation of the cosmological structures we observe today, the universe must necessarily have passed through a period of non-accelerating expansion, which means that the cosmological term must have decayed to a tiny value during some cosmological period in the past. The currently observed value (\ref{obs}), however, indicate that the universe is presently en\-tering an accelerated expansion era. Even though the reason why the universe is entering this accelerated expansion is an open question, the de Sitter special relativity leaves room for some speculations. For example, although the total energy $E = E_T - (1/4 l^2) E_K$ is conserved, the translational energy $E_T$ alone is not, in the sense that it can flow to $E_K$, and vice versa. As discussed above, at the initial instant of the universe, characterized by the huge cosmological term (\ref{PlanLam}), most of the energy was in the form of conformal energy $E_K$, and $E_T$ was very small. As the $\Lambda$ term decays and the universe expands, $E_K$ flows to $E_T$, and the entropy experience a continuous increase. This process can proceed until most of the energy is in the form of translational energy, and the cosmological term acquires a very small value.

Then comes the point: the existence of the conformal degrees of freedom allows the inverse process to occur without violating the energy conservation. In this process, $E_T$ flows back to $E_K$, entropy begins decreasing, and the value of $\Lambda$ would concomitantly grow up. In fact, considering that the entropy $S$ associated with the de Sitter horizon is given by \cite{gh} 
\be
S = \frac{k_B \, A_h}{4 \, l_P^2} \sim k_B \, \frac{l^2}{l_P^2},
\label{dsentro}
\ee
with $k_B$ the Boltzmann constant and $A_h = 4 \pi l^2$ the area of the horizon, a decrease in the entropy would imply a decrease in the de Sitter pseudo-radius $l$, and consequently an increase in the value of $\Lambda = 3/l^2$.\footnote{Notice that, in this context, the Boltzmann constant represents the entropy of a de Sitter spacetime with $l = l_P$. It can be considered, in this sense, a quantum of entropy \cite{grf}.} A possible source for this mechanism could be a loss of information occurring in black holes, which could result in a decrease of the universe entropy \cite{penrose0}. We reinforce that this mechanism turns out to possible just becausethere are two notions of energy, which can be transformed into each other. Of course, the details of these qualitative speculations have yet to be worked out, but they aren anyway useful for getting new insights on the physics underlying the de Sitter special relativity.

An interesting example of such new insight refers to the fact that an accelerated expansion does not mean necessarily that the universe is breaking apart, but rather that it is moving back towards its initial state, characterized by an infinite cosmological term. As we have already discussed, such state is represented by a flat, singular cone-spacetime, transitive under proper conformal transformations, in which all energy is in the form of proper conformal current (or equivalently, in the form of dark energy). According to the de Sitter relativity, therefore, in order to return to its initial state, the universe expansion does not need to stop, and then contracts back. Instead, it is through an accelerated expansion that the universe will be lead back to its initial state. We notice in passing that this picture points to a cyclic view of the universe, with the timelike future singular, conformal spacetime being identified with the initial state of the next era.

\section{Final remarks}

Cosmological observations and quantum gravity arguments seem to indicate the necessity of an invariant length-parameter at their respective scales. A simple and consistent way to introduce such parameter on both scales is arguably to replace ordinary special relativity by de Sitter special relativity. In fact, since the dispersion relation of this theory is invariant under change of scale of mass, energy and momentum, it can be applied at any scale, from quantum gravity to cosmology. The spatial isotropy and the relation between inertia-free frames in de Sitter relativity are still governed by the Lorentz group. Spacetime translations, however, are replaced by a combination of translations and proper conformal transformations. As a consequence, the notions of energy and momentum change, producing concomitant changes in the source of curvature, which turns out to be a combination of energy-momentum and proper conformal currents. Considering that a change in the local spacetime kinematics does not change general relativity, the energy-momentum tensor keeps its role of dynamic source of gravitation. The proper conformal current of ordinary matter, on the other hand, shows up as the kinematic source of the cosmological term $\Lambda$. The de Sitter special relativity, therefore, allows a new interpretation for the dark energy as an entity which is encoded in the structure of the kinematic group of spacetime, rather than produced by some exotic dynamical entity satisfying unphysical equations of state.

A crucial property of de Sitter special relativity is that, since neither energy-momentum nor proper conformal currents is individually conserved, energy-momentum can be transformed into proper conformal current, and vice-versa. Such mechanism yields a natural scenario for an evolving cosmological term. In fact, from a huge primordial $\Lambda$, in which most of the universe energy was in the form of dark energy, the cosmological term is then found to evolve according to the matter content of the universe, as given by the relation (\ref{dSitterEffects6}). After a decaying process, in order to allow the formation of the cosmological structures we see today, the universe has necessarily passed through a period of non-accelerating expansion, which means that the cosmological term must have assumed a tiny value during this period. Recent observations, however, indicate that the universe is presently entering an accelerated expansion era. This means that the energy-momentum current is transforming back into proper conformal current, which produces an increase in the value of the cosmological term. This mechanism is possible because, according to the de Sitter special relativity, an increasing $\Lambda$ makes the usual notions of time and space gradually fade away, until they completely disappear at the Planck scale, where only their conformal counterparts exist.

Now, as a consequence of the accelerated expansion, it is frequently argued that the universe would be driven either to a bleak future, a state that could be called cosmic loneliness, or even to a complete disintegration. However, as far as an increasing $\Lambda$ implies that matter and/or radiation is transforming into dark energy, such accelerated expansion means actually that the universe is moving back towards its initial state, characterized by an infinite cosmological term, in which all energy is in the form of dark energy. This picture points to a cyclic view for the universe, with the timelike future---represented here by the singular cone spacetime---identified with the initial state of the next era. According to this view, the universe would oscillate between a singular, conformally-transitive cone spacetime, characterized by in infinite cosmological term, and the translationally-transitive Minkowski spacetime, characterized by a vanishing cosmological term. These two extreme spacetimes are related by the inverse transformation
\be
x^\alpha \to - \, \frac{x^\alpha}{\sigma^2},
\ee
which is well known to lead the translation generators $\hat{P}_\alpha$ into the proper conformal generators $\hat{K}_\alpha$, and vice versa \cite{coleman}. In particular, due to the conformal invariance of the cone spacetime, which represents the singular border between two consecutive eras, this picture complies in many aspects with the conformal cyclic cosmology proposed by Penrose \cite{penrose}.

\section*{Acknowledgements}

The authors would like to thank FAPESP, CAPES and CNPq for partial financial support.


\end{document}